\documentclass{fmj2010}
\usepackage{graphicx}
\usepackage{natbib}
\usepackage{setspace}
\bibpunct{(}{)}{;}{a}{}{,}

\graphicspath{{/Users/cschang/data/mojave_sed/sed_plot/}}

\begin{document}

   \title{The Broadband Spectral Energy Distribution of the MOJAVE Sample}

   \author{C.~S.~Chang\inst{1}
          \and
                    E.~Ros\inst{2, 1}
           \and
                    M.~Kadler\inst{3, 4, 5} 
            \and
                     M.~F.~Aller\inst{6}
             \and
                     H.~D.~Aller\inst{6}
             \and
                    E.~Angelakis\inst{1}
              \and
                    L.~Fuhrmann\inst{1}   
               \and
                    I.~Nestoras\inst{1} 
               \and
                    H.~Ungerechts\inst{7}
          }

   \institute{Max-Planck-Institut f\"ur Radioastronomie, Auf dem H\"ugel 69, D-53121 Bonn, Germany
         \and
                       Departament d'Astronomia i Astrof\'isica, Universitat de Val\`encia, E-46100 Burjassot, Spain
          \and 
                       Dr. Remeis-Sternwarte \& ECAP, Sternwartstr. 7, D-96049 Bamberg, Germany
          \and
                       CRESST/NASA Goddard Space Flight Center, Greenbelt, MD 20771, USA
          \and
                       USRA, 10211 Wincopin Circle, Suite 500 Columbia, MD 21044, USA
          \and
                       Astronomy Department, University of Michigan, Ann Arbor, MI 48109-1042, USA
           \and
                       Institut de Radio Astronomie Millim\'etrique, Avenida Divina Pastora 7, Local 20, 18012 Granada, Spain
             }

   \abstract{We are constructing the broadband SED catalog of the MOJAVE sample from the radio to the $\gamma$-ray band using MOJAVE, \textit{Swift} UVOT/XRT/BAT, and \textit{Fermi}/LAT data, in order to understand the emission mechanism of extragalactic outflows and to investigate the site of high-energy emission in AGN. Since the launch of \textit{Fermi} $\gamma$-ray Space Telescope in August 2008, two thirds of the MOJAVE sources have been detected by \textit{Fermi}/LAT. Combining the results of high-resolution VLBI, X-ray, and $\gamma$-ray observations of the jet-dominated AGN sample, we want to pin down the origin of high-energy emission in relativistic jets. Here we present our overall project and preliminary results for 6 selected sources.  }

   \maketitle

\section{Introduction}
Blazars are extremely powerful objects which only represent a small
subset of active galactic nuclei (AGN), yet they dominate the
extragalactic radio and high-energy sky. Blazars are those AGN whose
jet is pointing towards us, which results in strong Doppler-boosting
of the emitted radiation. The small viewing angle is also responsible
for the superluminal jet motions observed in their jets. Since August
2008, the \textit{Fermi} Large Area Telescope (LAT) is performing a
continuous all-sky survey in $\gamma$-rays. After one year of
operation, \textit{Fermi}/LAT has detected 709 AGNs, of which 85\% are
blazars \citep{abdo10a}. The Monitoring Of Jets in Active galactic
nuclei with VLBA Experiments (MOJAVE) program has monitored a
radio-selected sample of 135 AGNs since the mid-1990s (J2000.0
declination\,$\leq-$20$^{\circ}$; galactic latitude
$|b|\leq$2.5$^{\circ}$; 15\,GHz VLBA flux density$\leq$1.5\,Jy), and
most of the MOJAVE sources are blazars due to the selection criteria
(see details in \citealt{lister09a, lister10}). In the one-year AGN
catalog of \textit{Fermi}, two-thirds of the MOJAVE sources are
detected \citep{boeck10}. Based on theoretical models, it is suggested
that the $\gamma$-ray and radio emission in blazars are closely
connected (\citealt{dermer93, sikora94}). \citet{dermer94} proposed
that the $\gamma$-ray emission from highly-beamed relativistic AGN
outflows originates near the base of the jet.  The high-resolution
capability of very-long-baseline interferometry (VLBI) enables us to
pin down the structure of extragalactic outflows up to the scale of
sub-parsecs, and to trace component ejections and evolution along
jets.

To understand the physical mechanisms ongoing in blazar jets, one of
the best approaches is to study the broadband spectral energy
distribution (SED). By using the multi-frequency data and
investigating the correlations, one can apply theoretical models to
the broadband emission to constrain further the parameters in the
local frame of AGN jets.

\section{The SED catalog of MOJAVE sources}
We want to investigate the broadband SED properties of the complete
radio-selected MOJAVE sample. All of the MOJAVE sources are X-ray
emitters \citep{kadler05}, and recent results also show that the
$\gamma$-ray brightness of AGN is correlated with VLBI jet properties
(\citealt{kovalev09, pushkarev09, savolainen10, lister09c, ojha10}).
A \textit{Swift} fill-in survey has been conducted since 2007 for the
MOJAVE sample (P.I.: M. Kadler), providing optical, UV, and X-ray
observations. As bright, well-studied radio sources, the MOJAVE
sources have very good flux-density sampling in the radio
band. Combining the \textit{Fermi}/LAT results, we are constructing
the broadband SED catalog from the radio to the $\gamma$-ray band of
the 135 MOJAVE sources of the statistical complete sample. Our goal is
to study their characteristics, and to determine the correlations of
the emission properties between different bands. In the MOJAVE sample,
there are 101 flat-spectrum radio quasars, 22 BL\,Lac objects, 8 radio
galaxies, and 4 unidentified AGNs. In the following sections, the
broadband SED data acquisition will be introduced. We will show
preliminary results of 6 selected sources which have good broadband
SED data coverage. All sources are in the \textit{Fermi} first-year
catalog, and were not included in the SED study of \textit{Fermi} LBAS
sample \citep{abdo10b}. We will discuss the preliminary results.

\section{Data acquisition and analysis}
Here we introduce the instruments and characteristics of the data
included in the MOJAVE SED catalog. In the $\gamma$-ray band, we
included the result of the 11-month \textit{Fermi}/LAT catalog
\citep{abdo10a}. We used the X-ray telescope (XRT) and burst alert
telescope (BAT) onboard \textit{Swift} to cover the X-ray band, and we
used the \textit{Swift} UV-optical telescope (UVOT) for UV and optical
band data. For each of the MOJAVE sources, we reduced one epoch of
\textit{Swift} XRT/UVOT data observed after August 2008 with the
longest XRT integration time. Many of the MOJAVE sources are weak in
X-rays, and typically we used epochs with XRT integration time around
5--10\,ksec. For the sources with no recent data, we submitted
\textit{Swift} Target-of-Opportunity (ToO) observing requests. As of
May 2010, only 5\% of the MOJAVE sources have no \textit{Swift} data
after August 2008, and are waiting observation.  Table
\ref{tab:source_parameter} summarizes our results of
\textit{Swift}/XRT data analysis, together with other relevant
parameters of the 6 sources.

\begin{table}[Ht]
\centering
\caption{Multi-wavelength facilities from the radio to the $\gamma$-ray bands that are used to construct the MOJAVE SED catalog.}
\resizebox{\columnwidth}{!}{
\begin{tabular}{@{}lll@{}}
\hline
\hline
\noalign{\smallskip}
Facility  &  Band   & Frequency (Hz)  \\
\noalign{\smallskip}
\hline
\noalign{\smallskip}
VLBA$^\mathrm{a}$   & Radio & 1.5$\times10^{10}$  \\
UMRAO  & Radio &  (4.8, 8, 14.5)$\times10^{9}$   \\
Effelsberg$^\mathrm{b}$  & Radio & (2.6,4.9,8.4,10.5, 14.6,23,32)$\times10^{9}$  \\
IRAM$^\mathrm{b}$  & Millimeter & (8.6,14.2,22.8)$\times10^{10}$  \\
\textit{Swift}/UVOT & UV-Optical & (5.5,6.9,8.6,11.3,13.4, 14.8)$\times10^{14}$  \\
\textit{Swift}/XRT & X-ray  & (7.25-242)$\times10^{16}$ \\
\textit{Swift}/BAT & X-ray & (3.6-36)$\times10^{18}$ \\
\textit{Fermi}/LAT & $\gamma$-ray & 4.8$\times10^{21}$\,to\,$>7.3\times10^{24}$ \\
\noalign{\smallskip}
\hline
\noalign{\smallskip}
\multicolumn{3}{@{}l@{}}{\footnotesize{$^\mathrm{a}$ MOJAVE program. }} \\
\multicolumn{3}{@{}l@{}}{\footnotesize{$^\mathrm{b}$ F-GAMMA project. }} \\
\end{tabular}	
\label{tab:mw_facility}
}
\end{table}

The 26-meter radio telescope in the University of Michigan Radio
Astronomy Observatory (UMRAO) has monitored a large number of AGN for
the past four decades \citep{aller85, aller03}, including all MOJAVE
sources. Here we included the UMRAO data closest in time to the
\textit{Swift} observation used.  We also included the Effelsberg
100\,m telescope and the IRAM 30\,m telescope observations in the
framework of the \textit{Fermi}-GST AGN Multi-frequency Monitoring
Alliance\footnote{The\,\,full\,\,F-GAMMA\,\,team,\,\,see
  \texttt{http://www.mpifr-bonn.mpg.de/div/vlbi/fgamma/teams.html}}
(F-GAMMA; \citealt{fulrmann10, angelakis10}). Also, we included the
broadband historical data from the NASA/IPAC Extragalactic
Database\footnote{\texttt{http://nedwww.ipac.caltech.edu/}}
(NED). Table \ref{tab:mw_facility} lists the facilities and wavebands
covered in our study.

\section{Results}
\begin{figure}
\centering
\includegraphics[clip,width=0.9\columnwidth]{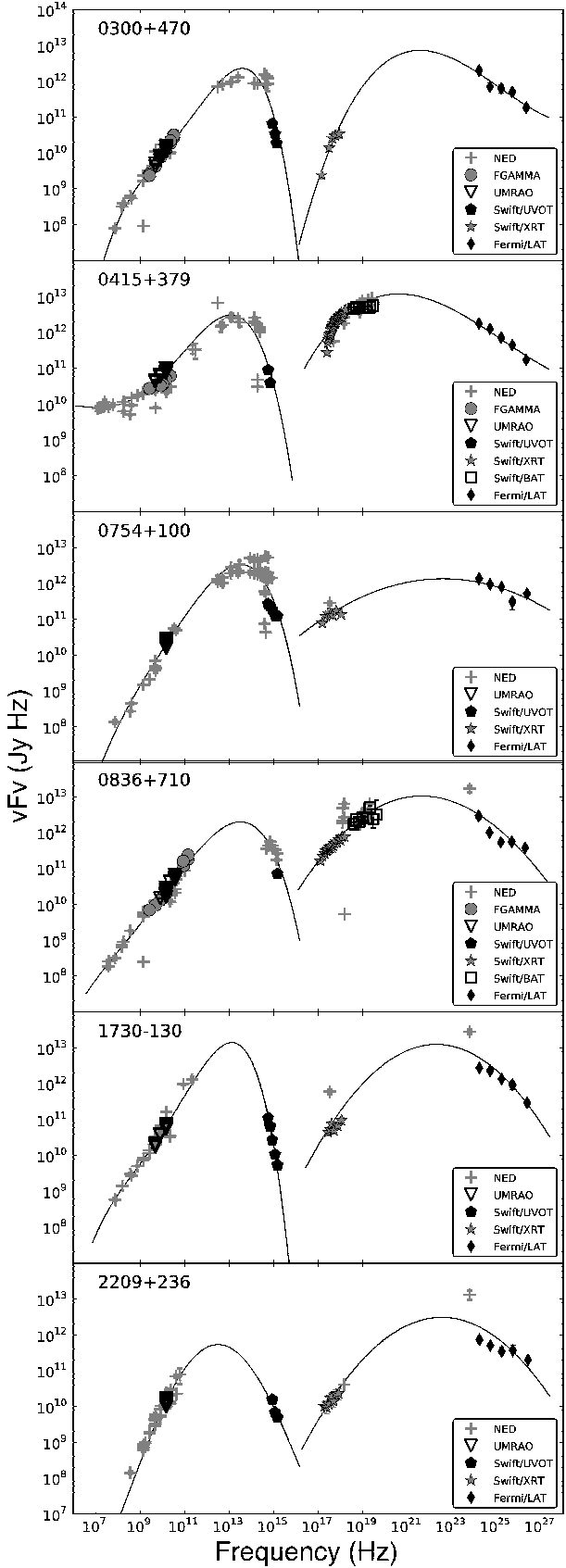}
 \caption{The broadband spectral energy distribution of 6 selected MOJAVE sources. Polynomial fitting results are shown as solid lines. }
 \label{fig:MOJAVE_SED}
\end{figure}

We selected 6 of the MOJAVE sources with good data coverage, and present the preliminary SED here (see Fig.~\ref{fig:MOJAVE_SED}). The presented sources are three flat-spectrum-radio quasars, two BL\,Lac objects, and one radio galaxy (see Table \ref{tab:source_parameter}). All of them are in the \textit{Fermi} first-year catalog, but were not included in the broadband SED study of \textit{Fermi} LBAS sample \citep{abdo10b}. 

The SEDs of the 6 MOJAVE sources show a classical double-peak blazar shape \citep{dermer93}. Current models suggest that the major part of emission from blazars are non-thermal, and the low-energy peak ($10^{7}-10^{16}$\,Hz in Fig.~\ref{fig:MOJAVE_SED}) is due to synchrotron emission from the radio jet, and the high-energy peak ($10^{17}-10^{27}$\,Hz in Fig.~\ref{fig:MOJAVE_SED}) can be interpreted as inverse Compton emission of various radiation sources, e.g., synchrotron self-Compton (SSC; \citealt{jones74, ghisellini89})  and external radiation Compton \citep{sikora94, dermer02}.  The physical processes involved in the SED study are complex, and the fact that we cannot distinguish different emission regions from a target source makes broadband SED modeling challenging. For example, the SED study of 48 \textit{Fermi} bright blazars \citep{abdo10b} shows that a homogeneous one-zone model with SSC mechanism cannot explain most of their results, and more complex models involving external Compton radiation or multiple SSC components are needed to model the blazar SEDs. In Fig.~\ref{fig:MOJAVE_SED}, we present the SED of the 6 sources in $\nu$-$\nu$F$_{\nu}$ plots. We performed a mathematical polynomial fit of 2nd to 4th order to the two humps, providing numerical values of the peak positions of the synchrotron and inverse-Compton components. Also, the total energy output of the low- and high- energy humps could be estimated from the polynomial fit. From Fig.~\ref{fig:MOJAVE_SED}, one can see that in each source, the total energy output of the low- and high- energy components differs. The high-energy hump of B0754$+100$ is significantly lower than the low-energy one, whereas the other sources have comparable height of the two humps. We discuss the 6 individual sources below.

\begin{table*}[Ht]
\caption{Summary of our X-ray and the $\gamma$-ray measurements of \citet{abdo10a} of the 6 sources in this study. }
\resizebox{\textwidth}{!}{%
\begin{tabular}{@{}cccccccccc@{}}
\hline
\hline
IAU  & 1LAC   & Source & z & $\beta_{\mathrm{app}}$  & F$_{\mathrm{X}}$ &  F$_{\gamma}$  & $\alpha_{\mathrm{X}}$  &  $\alpha_{\gamma}$   \\
Name  & 1FGL  &  type  &  &   & \scriptsize{[10$^{-13}$\,erg\,cm$^{-2}$\,s $^{-1}$]}  &  \scriptsize{[10$^{-13}$ photon\,\,\,\,\,\,\,\,\,}  &  &  \\
       &        &         &       &         &        & \scriptsize{MeV$^{-1}$\,cm$^{-2}$\,s$^{-1}$]} & & \\
(1)  &  (2) &  (3)  & (4) & (5)   & (6)  & (7)  & (8)  & (9) \\
\hline
B0300$+$470$^{*}$  & J0303.1$+$4711 & B  & -           & -                         & 7.38  &  49.9$\pm$7.9   & 1.38$\pm$0.41 & 2.56$\pm$0.13 \\
B0415$+$379  & J0419.0$+$3811          & G & 0.049   &  5.9$\pm$0.1 & 454  &   76.3$\pm$15.8 & 1.73$\pm$0.01 & 2.61$\pm$0.16 \\
B0754$+$100  & J0757.2$+$0956          & B  & 0.266   & 14.4$\pm$1.2    & 29.8  &  75.2$\pm$6.7  & 1.81$\pm$0.08 & 2.39$\pm$0.08 \\
B0836$+$710  & J0842.2$+$7054          & Q & 2.218   & 25.4$\pm$1.0    & 122  &   366$\pm$32    & 1.43$\pm$0.02 & 2.98$\pm$0.12 \\
B1730$-$130   & J1733.0$-$1308           & Q  & 0.902   & 35.7$\pm$2.1  &  12.1  &  85.4$\pm$7.6  &  1.98$\pm$0.33 & 2.34$\pm$0.07 \\
B2209$+$236  &  J2212.1+2358             & Q  & 1.125   & 3.4$\pm$0.5 & 4.80  &  3.3$\pm$ 0.7& 1.43$\pm$0.60 &  2.13$\pm$0.19   \\
\hline
\multicolumn{9}{@{}l@{}}{\footnotesize{(1) Source Name; (2) The first LAT AGN catalog name \citep{abdo10a}; (3) Optical class (B: BL\,Lac, G: radio galaxy, Q: quasar); }} \\
\multicolumn{9}{@{}l@{}}{\footnotesize{(4) Redshift; (5) The maximum apparent projected speed measured by the MOJAVE team \citep{lister09b}; (6) X-ray flux }} \\
\multicolumn{9}{@{}l@{}}{\footnotesize{measured by \textit{Swift}/XRT (0.2-10\,keV); (7) $\gamma$-ray flux measured by \textit{Fermi}/LAT using 11-month data \citep{abdo10a}; }} \\
\multicolumn{9}{@{}l@{}}{\footnotesize{(8) X-ray photon index; (9) $\gamma$-ray photon index.}} \\
\multicolumn{9}{@{}l@{}}{\footnotesize{$^{*}$ Luminosity distance unknown, measured maximum apparent speed 287$\pm$23 $\mu$as yr$^{-1}$.}} \\
\end{tabular}	
\label{tab:source_parameter}
}
\end{table*}

\paragraph{B0300$+$470}
4C\,+47.08 is a BL\,Lac object, which shows a compact asymmetric morphology in VLBI observations \citep{lister09a}. It displays a one-sided halo at kilo-parsec scale. This source is variable  on a monthly timescale at centimeter wavelengths \citep{aller85} in total flux and linear polarization, and the source is core-dominated \citep{nan99}. We analyzed one epoch of \textit{Swift} observations of this source obtained in September 2008 (ID 00036235005) with an integration time of 7\,ks. 

\paragraph{B0415$+$379}
3C\,111 is a well-studied broad-line radio galaxy which shows a classical Fanaroff and Riley Class II morphology on kiloparsec scales \citep{linfield84}. It hosts a highly collimated one-sided jet emitting from the central core to the northeastern lobe (see e.g., \citealt{kadler08}). 3C\,111 is associated with the EGRET source 3EG J0416$+$3650,  and the broadband SED study of the historical data suggests that the SED profile is similar to EGRET flat spectrum radio quasars \citep{hartman08}.  Here we analyzed \textit{Swift} data observed in January 2010 (obsid 00036367005) with an integration time of 9\,ks.   

\paragraph{B0754$+$100}
B0754$+$100 is cataloged as a low-frequency peaked BL\,Lac (LBL) by \citet{fiorucci04}. By the blazar classification of \citet{abdo10b}, B0754$+$100 is a low synchrotron peaked blazar ($\nu_{\mathrm{peak}}\leq10^{14}$\,Hz) based on our fitting result.  VLA observations showed that there is a diffuse halo around the core at 1.5 and 5\,GHz up to scales of  $>$120 kpc \citep{antonucci85, kollgaard92}. Recently, B0754$+$100 was reported to be flaring in the near-infrared band \citep{carrasco10}. We requested \textit{Swift} ToO request for this source, which was observed in February 2010 (obsid 00036195002) with an integration time of 9\,ks.

\paragraph{B0836$+$710}
4C\,+71.07 is a luminous quasar which hosts a radio jet extending up to kiloparsec scales. Broadband variability was observed in this source \citep{otterbein98}, and space VLBI observations have revealed detailed jet structures \citep{lobanov98b, perucho07}. Here we included the \textit{Swift} data of B0836$+710$ observed in February 2009 with an integration time of 9\,ks (obsid 00036376005).  

\paragraph{B1730$-$130}
NRAO 530 is a high-polarized radio quasar which hosts a double-sided kilo-parsec scale jet. This source is actively variable in the radio \citep{marscher81, feng06}, optical \citep{pollock79}, X-ray \citep{foschini06}, and $\gamma$-ray \citep{hartman99} wavebands. We used the \textit{Swift} data of NRAO 530 observed in June 2009 with an integration time of 5\,ks (obsid 00035387012).   

\paragraph{B2209$+$236}
B2209$+$236 is a core-dominated flat-spectrum radio quasar. The radio turn-over frequency of this source is above 5\,GHz \citep{dallacasa00}, which indicates that the source is very compact and is possibly young.  Here we used the \textit{Swift} observation obtained in April 2009  with an integration time of 9\,ks in the SED study (obsid 00036359002).

\section{Outlook}  
As of May 2010, we completed the data collection and construction of quasi-simultaneous SEDs during the time between August 2008 to July 2009. Almost all sources have nearly simultaneous observations from radio to X-ray with good data coverage. We are compiling the statistical properties of the 135 sources (e.g., correlation study between VLBI properties, X-ray and $\gamma$-ray parameters, and broadband SED characteristics), and the results will be presented elsewhere (Chang et al., in preparation). Physical SED models will be applied to the whole sample, to shed light on the mechanisms of general AGN picture.

\begin{acknowledgements}
We thank especially M. B\"ock, L. Barrag\'an, J. Wilms, C. M. Fromm, and C. Ricci for valuable discussions. This research was supported by the EU Framework 6 Marie Curie Early Stage Training program under contract number MEST/CT/2005/19669 ESTRELA. CSC is a member of the International Max Planck Research School for Astronomy and Astrophysics. This research includes data from observations with the 100-m telescope of the MPIfR at Effelsberg. This research has made use of data from the MOJAVE database that is maintained by the MOJAVE team \citep{lister09a}, and the University of Michigan Radio Astronomy Observatory which has been supported by the University of Michigan \citep{aller03}.
\end{acknowledgements}

\begin{spacing}{0.87}
\bibliographystyle{aa}
\bibliography{CSChang}
\end{spacing}

\end{document}